\documentclass[10pt, a4paper]{article}
\usepackage{amsmath}
\usepackage[utf8]{inputenc}
\usepackage[T1]{fontenc}
\usepackage{times}
\usepackage{graphicx}
\usepackage{amsfonts}

\usepackage{geometry}
\newgeometry{tmargin=2cm, bmargin=2cm, lmargin=2cm, rmargin=2cm}

\newcommand{\sph}[1]{\left\langle #1 \right\rangle}

\begin{document}

\title{Is the motion of a single SPH particle droplet/solid physically correct?}
\author{Kamil Szewc$^1$, Katarzyna Walczewska-Szewc$^2$, Michal Olejnik$^1$ \\ $^1$IFFM, Polish Academy of Sciences, ul. Fiszera 14, 80-231 Gda\'nsk, Poland\\
$^2$In-Silico, Numerical Laboratory, ul. Jodowa 1B/5, 80-680 Gda\'nsk, Poland}

\maketitle

%\cgsn{National Science Centre, Poland}{\emph{Opus 6} no DEC-2013/11/B/ST8/03818}

\begin{abstract}
In recent years the Smoothed Particle Hydrodynamics (SPH) approach gained popularity in modeling multiphase and free-surface flows. In many situations, due to certain reasons, interface and free-surface fragmentation occurs. As a result single SPH particle solids/droplets of one phase can appear and travel through other phases. In this paper we investigate this issue focusing on a movement of such single SPH particles. The main questions we try to answer here are: is movement of such particles physically correct? What is its physical size? How numerical parameters affect on it? With this in mind we performed simple simulations of solid particles falling due to gravity in a fluid. Considering three different diameters of a single particle, we compared values of the drag coefficient and the velocity obtained through the SPH approach with the experimental and the analytical reference data. In the way to accurately model multiphase flows with free-surfaces we proposed and validated a novel SPH formulation.   
\end{abstract}

\maketitle

\vspace{-6pt}

\section{Introduction}
The Smoothed Particle Hydrodynamics method (SPH) is a mesh-free, particle-based approach for fluid-flow modelling. In the early stage it was developed to simulate some astrophysical phenomena at the hydrodynamic level~\cite{Monaghan 1992}. The main idea behind the SPH method is to introduce kernel interpolants for flow quantities in order to represent fluid dynamics by a set of particle evolution equations. Due to its Lagrangian nature, for multiphase flows, there is no necessity to handle (reconstruct or track) the interface shape as in the grid-based methods. Therefore, there is no additional numerical diffusion related to the interface handling. For this reason and the fact that the SPH approach is well suited to problems with large density contrast, free-surfaces and complex geometries, the SPH method is increasingly used for hydro-engineering and geophysical applications, for review see~\cite{Monaghan 2012, Violeau 2012}. 

In many situations, due to both: physical processes and numerical errors, the interface in multi-phase flows and the surface in free-surface flows can be fragmented, see~\cite{Monaghan 2012, Colagrossi and Landrini 2003, Grenier et al. 2009, Szewc et al. 2015, Szewc et al. 2013}. As a result, the single SPH particle solids/droplets (or even bubbles) can appear and move in the opposite phase or travel without any friction through the regions without particles (representing gaseous phase in free-surface flows).  
Since the reason of the appearance of single SPH particle solids/droplets may be different depending on the nature and scale of the problem, we do not intent to discuss the causes in detail. 
However, in some papers, eg. \cite{Wieth et al. 2015a, Wieth et al. 2015b}, the authors model a dispersed phase (droplets) moving in the other liquid as a single or a small number of the SPH particles.
In the present paper, we focus on the problem of physical correctness of such single particle solids/droplets motion. We try to answer the following questions:
is the motion of a single particle droplet physically correct? What is such a droplet's diameter? What is an influence of the SPH numerical parameters such as $h$, $h/\Delta r$ or the shape of the kernel on the motion of the single particle?

In the following considerations we assume that the reader is familiar with the SPH basics. 
Those not familiar with the method may find useful information on the SPHERIC - SPH European Research Interest Community web page~\cite{SPHERIC}. Another reliable source of information is a handbook by \emph{Violeau (2012)}~\cite{Violeau 2012}.

\section{Multiphase SPH formulation}
The full set of governing equations for incompressible viscous flows is composed of the Navier-Stokes (N-S) equation
\begin{equation}
    \frac{d\mathbf u}{dt} = -\frac{1}{\varrho} \nabla p + \frac{1}{\varrho} \left(\nabla \mu \cdot \nabla \right) \mathbf u + \mathbf f,
\end{equation}
where $\varrho$ is the density, $\mathbf u$ velocity, $t$ time, $p$ pressure, $\mu$ the dynamic viscosity and $\mathbf f$ an acceleration; the continuity equation
\begin{equation}
    \frac{d\varrho}{dt} = -\varrho\nabla \cdot \mathbf u \xrightarrow{\varrho=const} \nabla \cdot \mathbf u = 0,
\end{equation}
and the advection equation (Lagrangian formalism)
\begin{equation}
    \frac{d\mathbf r}{dt} = \mathbf u,
\end{equation}
where $\mathbf r$ denotes position of fluid element.

The governing equations can be expressed in the SPH formalism in many different ways. In general, two SPH approximations: integral interpolation and discretization, lead to the the basic SPH relationship
\begin{equation} \label{basic SPH}
    \sph{A}(\mathbf r) = \sum_b A(\mathbf r_b) W(\mathbf r - \mathbf r_b, h) \Omega_b,
\end{equation}
where $A$ is a physical field (in a sake of simplicity we consider a scalar field only), $W$ is a weighting function (kernel) with parameter $h$ called the smoothing length, while $\Omega$ is the volume of the SPH particle. In the present paper we use the Wendland kernel~\cite{Wendland 1995} in the form
\begin{equation} \label{quintic Wendland}
W(\mathbf r, h) = C  \left \{
 \begin{array}{cl}
  \left( 1-q/2 \right)^4 (2q+1) &  \text{for} \;\; q \leq 2, \\
  0                                     &  \text{otherwise},     \\
  \end{array}
  \right.
\end{equation}
where $q = |\mathbf r|/h$ and the normalization constant is $C=7/4\pi h^2$ (in 2D) 
or $21 / 16\pi h^3$ (in 3D). For details how the choice of the kernel and smoothing length affect results see~\cite{Szewc et al. 2012a}. It is important to note here that the SPH basic approximation, Eq.~(\ref{basic SPH}), is common also in other numerical particles-based approaches, e.g. Moving Particle Semi-implicit Method (MPS)~\cite{Koshizuka et al. 1998}. The SPH method differs from other methods in aspect of approximation of differentiation operator. Assuming the kernel symmetry, nabla operator can be shifted from the action on the physical field to the kernel
\begin{equation}
    \sph{\nabla A}(\mathbf r) = \sum_b A(\mathbf r_b) \nabla W(\mathbf r - \mathbf r_b, h) \Omega_b.
\end{equation}

It is important to note that although different SPH formulations can be obtained from the same governing equations, some of them may not by applicable for certain types of flows. For instance, the most common SPH form of the N-S pressure term
\begin{equation}
    \sph{\frac{\nabla p}{\varrho}}_a = -\sum_b m_b \left(\frac{p_a}{\varrho_a^2} + \frac{p_b}{\varrho_b^2} \right) \nabla_a W_{ab},
\end{equation}
where $\nabla_a W_{ab}=\nabla_a W(\mathbf r_a - \mathbf r_b, h)$, is very useful for modelling single-phase flows with free-surfaces. However, when applied for multi-phase flows with interfaces it leads to some instabilities at interface, for details see \cite{Colagrossi and Landrini 2003, Szewc et al. 2012b}. To the best of our knowledge, there are three SPH formulations constructed to model multi-phase flows: \emph{Colagrossi and Landrini (2003)}~\cite{Colagrossi and Landrini 2003}, \emph{Hu and Adams (2006)}~\cite{Hu and Adams 2006} and \emph{Grenier et al. (2009)}~\cite{Grenier et al. 2009}. For the purposes of this work, we decided to use the \emph{Hu and Adams (2006)} formalism. In this approach, the N-S pressure term becomes
\begin{equation} \label{H-A pressure term}
    \sph{\frac{\nabla p}{\varrho}}_a = -\frac{1}{m_a}\sum_b \left(\frac{p_a}{\Theta_a^2} + \frac{p_b}{\Theta_b^2} \right) \nabla_a W_{ab},
\end{equation}
where $\Theta$  is inverse of particle volume, and the viscous N-S term is
\begin{equation} \label{H-A viscous term}
    \sph{\left(\nabla \mu \cdot \nabla \right) \mathbf u}_a = \frac{1}{m_a} \sum_b \frac{2 \mu_a \mu_b}{\mu_a + \mu_b} \left( \frac{1}{\Theta_a^2} + \frac{1}{\Theta_b^2} \right) \frac{\mathbf r_{ab} \cdot \nabla_a W_{ab}}{r_{ab}^2 + \eta^2} \mathbf u_{ab},
\end{equation}
where $\eta=0.01h$ is a small regularizing parameter, while the continuity equation has the form (of the density definition)
\begin{equation} \label{H-A continuity equation}
    \varrho_a = m_a \sum_b W_{ab} = m_a \Theta_a.
\end{equation}
In this variant of the continuity equation the density field is represented only by a spatial distribution of neighbouring particles, but not by their masses. Therefore, in multi-phase flows, particles located near an interface but belonging to different phases may interact without having their density affected by the other fluid. The main drawback of Eq.~(\ref{H-A continuity equation}) is inability to deal with free-surfaces. Lack of the SPH particles apart from the fluid causes the underestimation of the density field near the free-surfaces. 
To overcome this problem, we decided to take a variation of Eq.~(\ref{H-A continuity equation}) (assuming constant mass of particles)
\begin{equation} \label{variation of H-A continuity equation}
    \delta \varrho_a = m_a \sum_b \left(\delta \mathbf r_a - \delta \mathbf r_b \right) \cdot \nabla_a W_{ab}.
\end{equation}
The new variant of the continuity equation can be explicitly obtained from Eq.~(\ref{variation of H-A continuity equation}) by differentiation with respect to time
\begin{equation} \label{S-O continuity equation}
    \frac{d\varrho_a}{dt} = m_a \sum_b \mathbf u_{ab} \cdot \nabla_a W_{ab}.
\end{equation}
It is important to note that Eqs.~(\ref{S-O continuity equation}) and (\ref{H-A pressure term}) are variationally consistent, for details see \cite{Grenier et al. 2009, Monaghan 2005}.
The definite advantage in using Eq.~(\ref{S-O continuity equation}) over (\ref{H-A continuity equation}) is that the density only varies when particles move relative to each other. Therefore, despite a decrease of number of particles near the free-surface, this does not affect significantly the calculation of the density field (except the accuracy -- smaller number of particles within the kernel range). The important difference between the \emph{Hu and Adams (2006)} formulation and our approach is that we do not calculate $\Theta_a$ by summation over neighbouring particles, cf.~Eq.~(\ref{H-A continuity equation}). Instead, in Eqs.~(\ref{H-A pressure term}) and (\ref{H-A viscous term}), we explicitly write $\Theta_a = \varrho_a / m_a$.

In the present work, we decided to use the most common method of implementing the incompresibility -- Weakly Compressible SPH (WCSPH). It involves the set of governing equations closed by a suitably-chosen, artificial equation of state, $p=p(\varrho)$. Following the mainstream, we decided to use the Tait equation of state
\begin{equation}
    p = \frac{c^2 \varrho_0}{\gamma} \left[\left( \frac{\varrho}{\varrho_0} \right)^\gamma - 1 \right],
\end{equation}
where $\varrho_0$ is the initial density. The sound speed $c$ and a parameter $\gamma$ are suitably chosen to reduce the density fluctuations down to $1\%$. In the present work we set $\gamma=7$ and $c$ at the level at least 10 times higher than the maximal fluid velocity. It is worth noting that two alternative incompressibility treatments exists: Incompressible SPH (ISPH) where the incompressibility constaint is explicity enforced through the pressure correction procedure to satisfy $\nabla \cdot \mathbf u=0$~\cite{Szewc et al. 2012a, Cummins and Rudman 1999, Hu and Adams 2007, Shao and Lo 2003, Xu et al. 2009} and Godunov SPH (GSPH) where the acoustic Riemann solver is used~\cite{Rafiee et al. 2012}.
The boundary conditions are fulfilled applying the ghost-particle method~\cite{Szewc et al. 2012a, Cummins and Rudman 1999}.

In order to validate the proposed formulation, we decided to model the Rayleigh-Taylor instability, which is one of the generic multi-phase tests. Our case involves two immiscible fluids enclosed in a rectangular domain of the edges' sizes: $L_x=L$ and $L_y=2L$. Initially, the phases are separated by the interface located at $y/L-1-0.15\sin{(2\pi x/L)}$. The density and viscosity ratios between the upper (U) and the lower (L) phases are: $\varrho_U/\varrho_L=1.8$ and $\nu_U/\nu_L=1$ respectively. Since the system is subject to gravity, $g$, and there is no surface tension, the system destabilizes resulting in generation of vorticity. The Reynolds number of the considered case is $Re=g^{1/2} L^{3/2} \nu_L^{-1} = 420$. 
\begin{figure}
	\centering
    \includegraphics[width=0.99\textwidth]{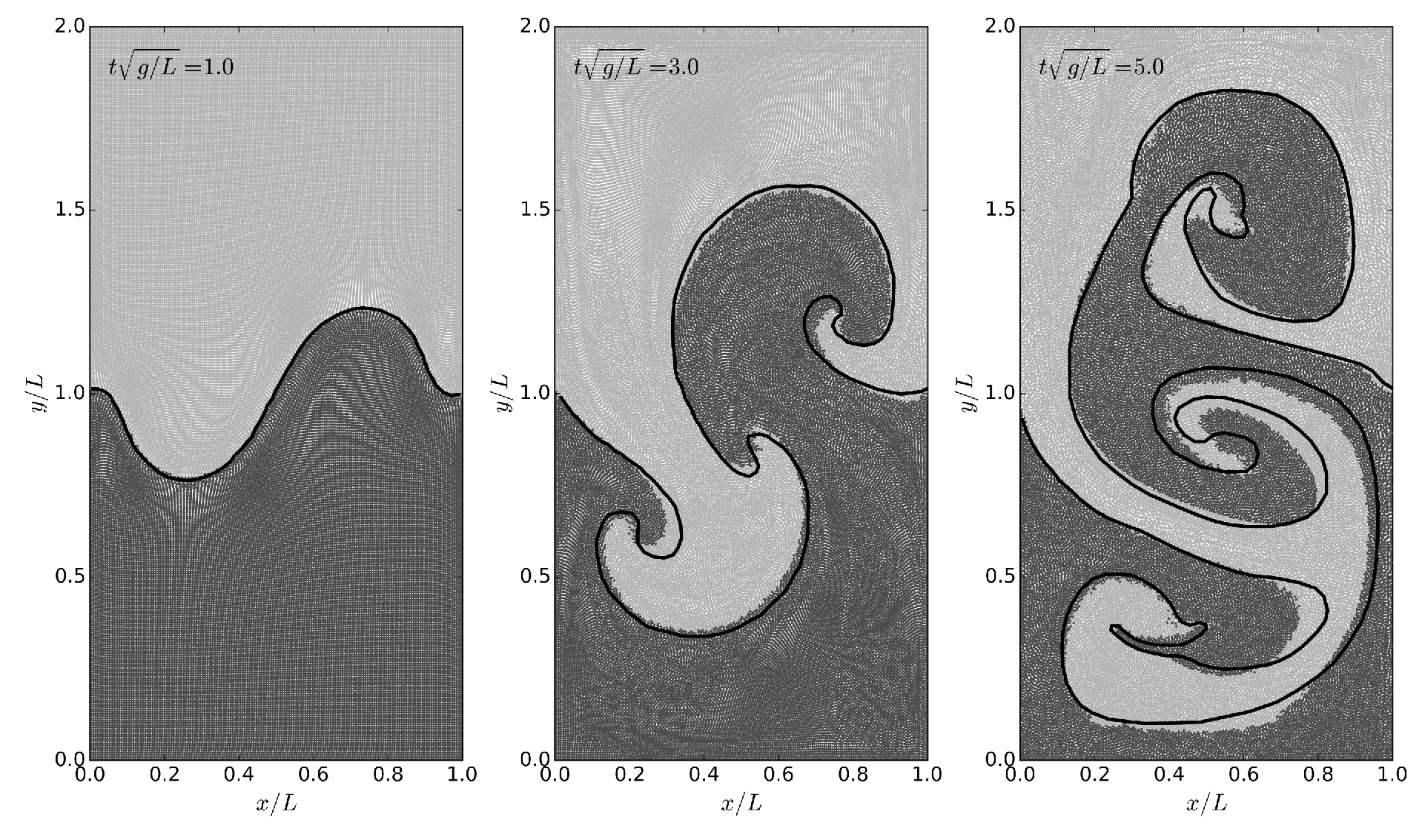}
    \caption{The Rayleigh-Taylor instability ($Re=420$) calculated using the proposed SPH formalism ($h/L=1/64$, $h/\Delta r=2.5$); black line denotes the interface position calculated using the Level-Set method ($312\times 624$ cells) by \emph{Grenier et al. (2009)}~\cite{Grenier et al. 2009}.}
    \label{fig:rt}
\end{figure}
The simulations were performed for $h/L=1/64$, $h/\Delta r=2.5$ and the Wendland kernel. Figure~\ref{fig:rt} presents particle positions at $t(g/L)^{1/2}=1$, $3$ and $5$ compared with the reference solutions from the Level-Set method ($312\times 624$ cells) computed by \emph{Grenier et al. (2009)}~\cite{Grenier et al. 2009} (black line). The results obtained with SPH show good agreement with the reference data.

\section{Single particle droplets/solids}
\begin{figure}
	\centering
    \includegraphics[width=0.99\textwidth]{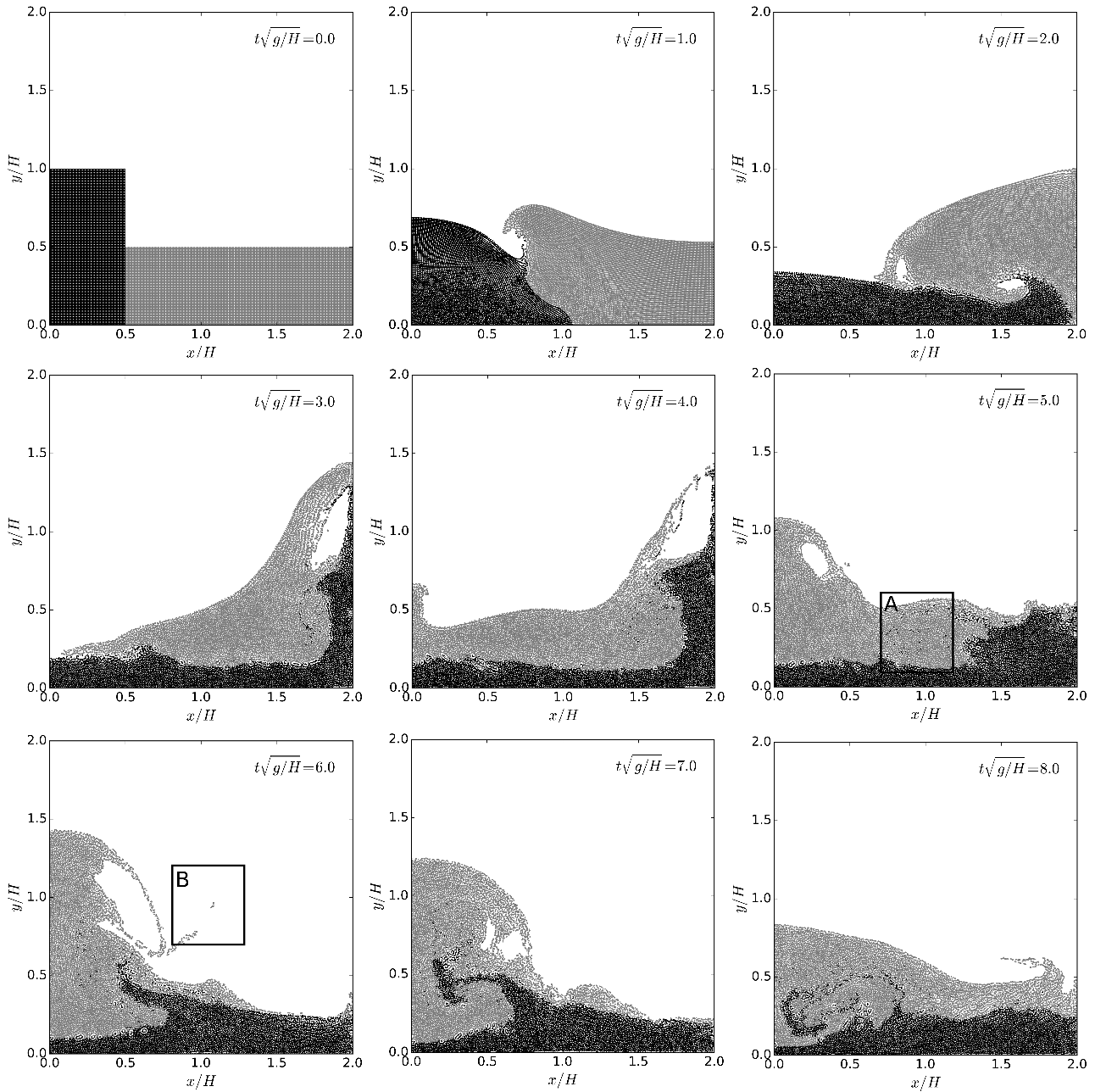}
    \caption{Evolution of the dam breaking test case $(Re=2000)$ calculated using the SPH approach ($h/H=1/32$, $h/\Delta r=2$); frames A and B (zoomed in Fig.~\ref{fig:damAB}) indicate regions where some single SPH particle droplets/solids appears and travel through the domain.}
    \label{fig:dam}
\end{figure}

\begin{figure}
	\centering
	\begin{tabular}{cc}
    	\includegraphics[width=0.49\textwidth]{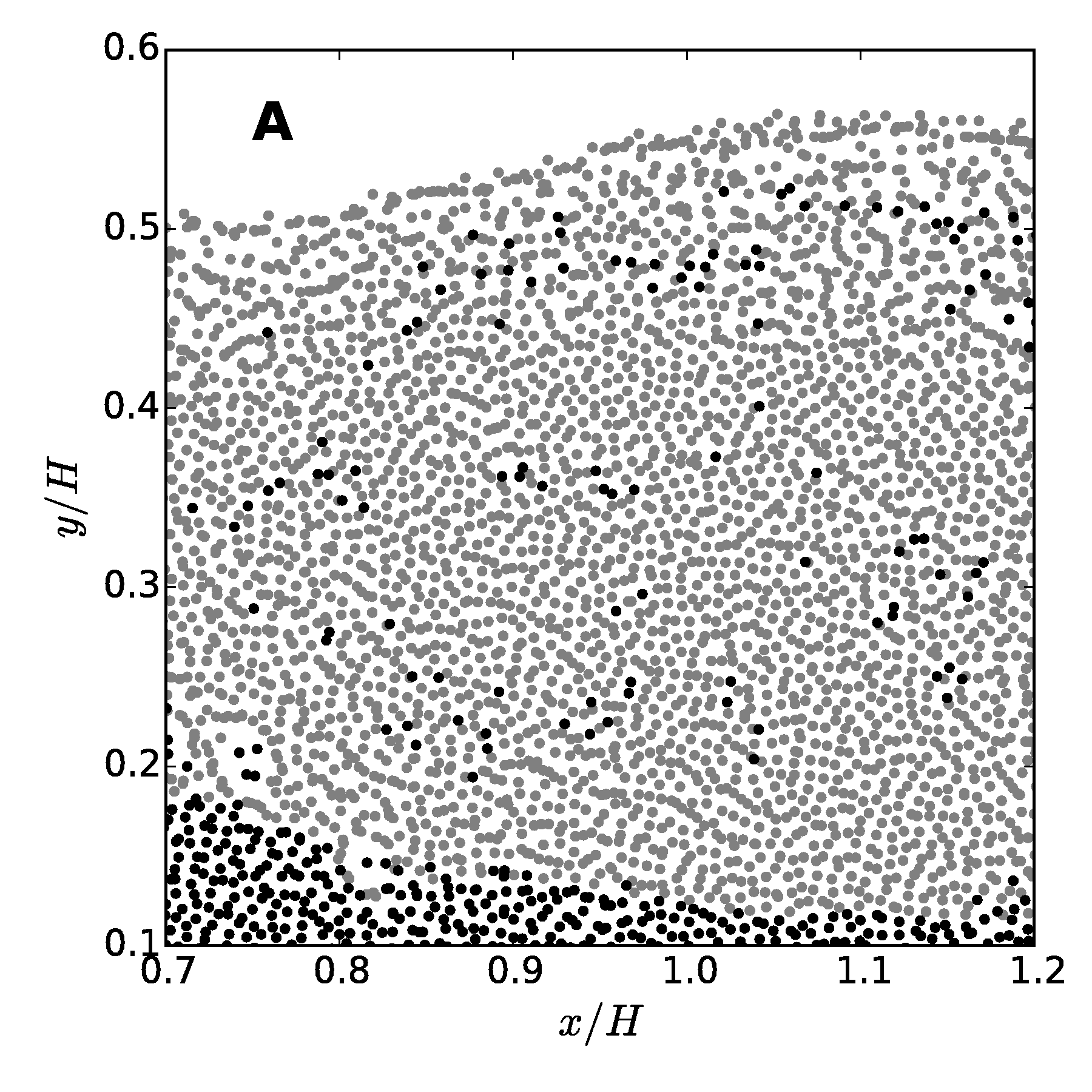} &
    	\includegraphics[width=0.49\textwidth]{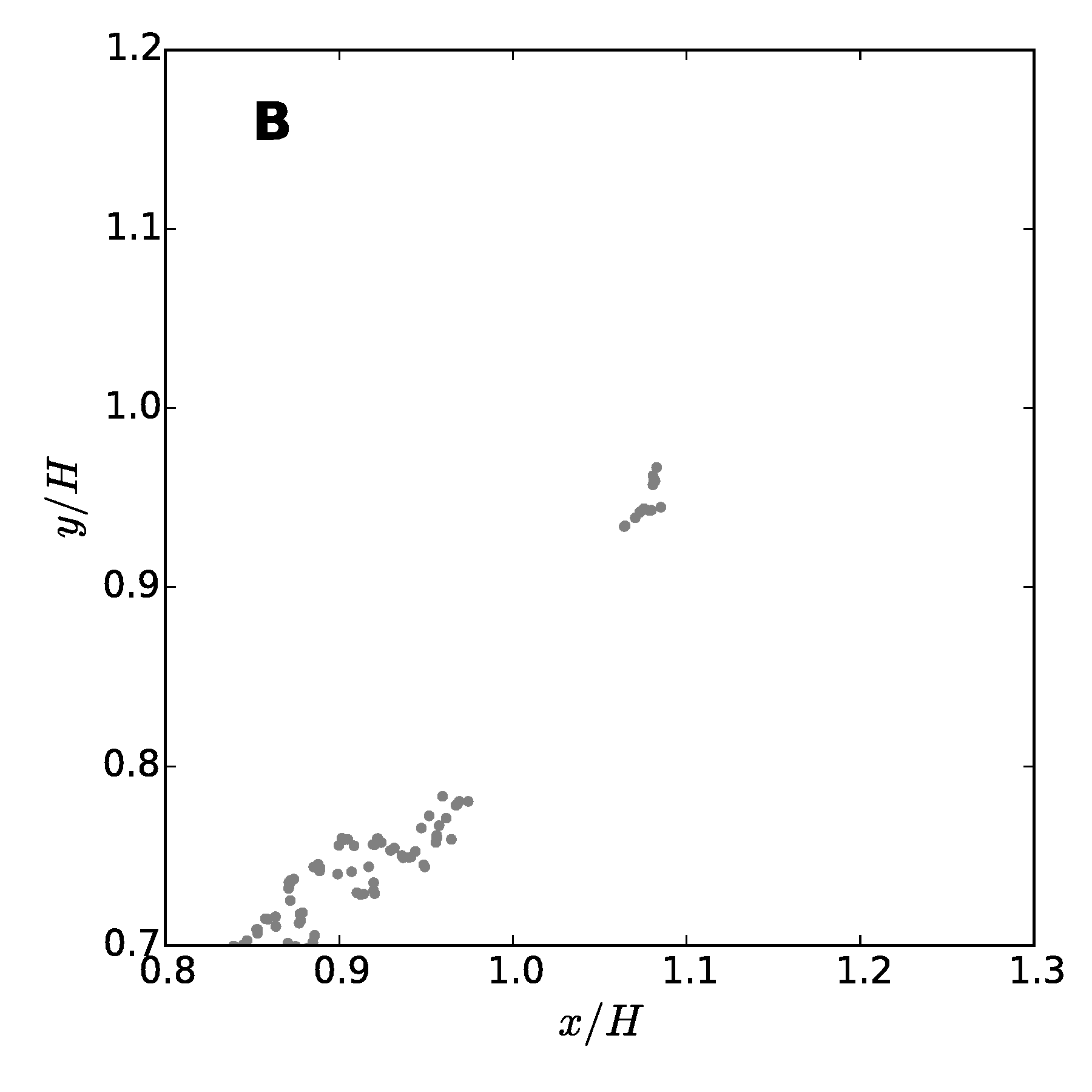} 
    \end{tabular}
    \caption{Zoom of the regions where single SPH particles droplets/solids appear and travel in the dam breaking test case simulation; frame A shows swarms of single particles of the denser phase within the lighter one; frame B presents the fragmentation of the free-surface.}
    \label{fig:damAB}
\end{figure}

To illustrate interface fragmentation problem, we consider the multi-phase dam-breaking problem whose evolution is presented in Fig.~\ref{fig:dam}. This simulation involves two liquid columns: black ($B$) and gray ($G$) enclosed in the square box of the edge size $2H$, where $H$ is the heigh of the black column. The density and viscosity ratios between phases are respectively $\varrho_B/\varrho_G=4$ and $\mu_B/\mu_G=2$. The Reynolds number is $Re=\varrho_B g^{1/2}H^{3/2}\mu_B^{-1}=2000$. The simulations were performed for $h/H=1/32$, $h/\Delta r=2$ and the Wendland kernel. Due to the gravity, $g$, the denser (black) fluid spreads over the bottom of the tank. Since the tongue of black phase propagates, it induces a very violent sloshing flow of the lighter (gray) phase. The presented simulation reveals two events of interface fragmentation whose physical interpretation is not clear. The first one, indicated in Fig.~\ref{fig:dam} by frame A and zoomed in Fig.~\ref{fig:damAB}(A) manifests itself as the appearance of the single particle droplets of the denser phase within the lighter one. The second one, indicated in Fig.~\ref{fig:dam} by frame B and zoomed in Fig.~\ref{fig:damAB}(B) results in fragmentation of the free-surface what leads to the release of particle droplets into the area free of particles, which represents an air.
From the physical point of view, in both: A and B cases, the single droplets could appear in the flow as an effect of the instabilities in the flow, e.g. Rayleigh-Taylor or Kelvin-Helmholtz in the case A, the Rayleigh-Plateau in the case B, and shear. However, in the case of the SPH modelling, the size of the single droplet is determined by the size of a single SPH particle. What is more, since the single SPH particle is represented by position (point), density and mass, the definition of the SPH particle size is not obvious. Although the mass of particle is usually fixed, the density can change which implies the change of the particle volume. Therefore, the physical interpretation of the process of appearance of single droplets is not evident. Probably, in most of the cases such a behaviour has to be interpreted as a numerical error related to the low-resolution, sub-kernel particle motion, or/and non-physical pressure pulsation in WCSPH or lack of volume conservation in ISPH. The problem is that the dense droplet swarms as presented in Fig.~\ref{fig:damAB}(A) or free-falling droplets, see Fig.~\ref{fig:damAB}(B), hitting (with high velocity) the free-surface can significantly modify the flow in the whole domain. Nevertheless, for further discussion, let us assume that the resolution is high enough to interpret appearance of the single particle droplet as a physically correct process. Is the motion of a single particle droplet physically correct? What is a diameter of such a droplet? What is an influence of the SPH numerical parameters such as $h$, $h/\Delta r$ or the shape of the kernel on the motion of the single particle?

\section{Drag coefficient of single particle droplet/solid}
\begin{figure}
	\centering
    \includegraphics[width=0.99\textwidth]{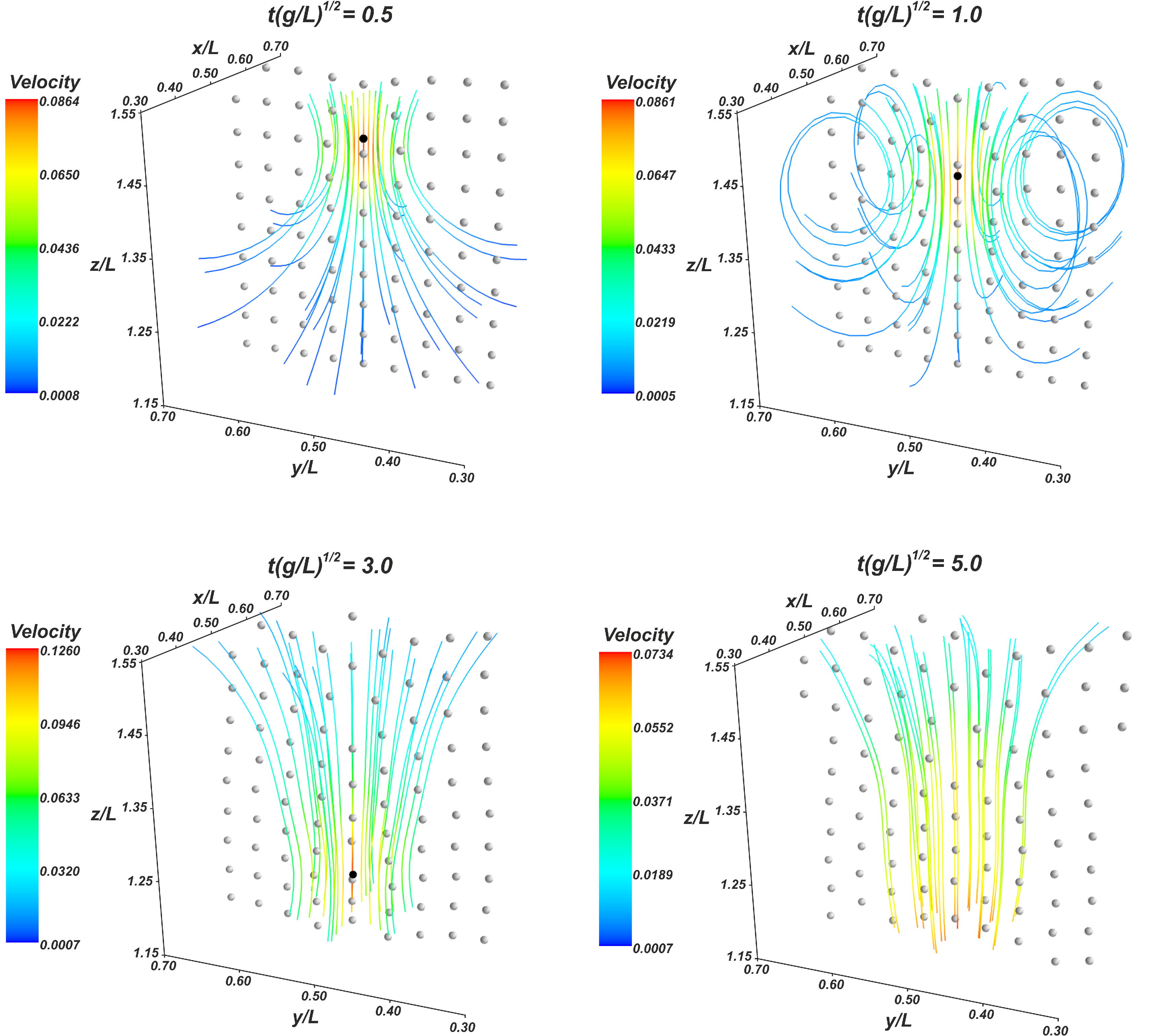}
    \caption{Results of the SPH modeling of single SPH particle solid (black one) falling down due to gravity; gray particles represent fluid; the streamlines are calculated using the Mayavi~\cite{Mayavi} plotting tool.}
    \label{fig:particle}
\end{figure}

In order to answer all raised questions about the motion of single droplet, we decided to perform a simple numerical experiment. We considered a cuboid of the edges' sizes: $L_x=L$, $L_y=L$ and $L_z=2L$. The whole domain is filled with the fluid of density, $\varrho_L$. Since the hydrostatic pressure balances the effect of the gravitational acceleration, the system is at rest. At some point, we change the density of one particle (located centrally in the upper part of the domain) into the density of the solid/droplet particle, $\varrho_S$. Due to the gravity, the solid particle begins to fall. 
It accelerates to reach the terminal velocity $u_T$. The evolution of the described set-up is presented in Fig.~\ref{fig:particle}.
In reality, the solid object inserted into the liquid should move according to the relation
\begin{equation} \label{general equation of motion}
	m \frac{d\mathbf u}{dt} = \mathbf F_G + \mathbf F_B + \mathbf F_D,
\end{equation}
where $\mathbf F_G=m \mathbf g$ is the gravitational force, $\mathbf F_B$ is the buoyancy force and $\mathbf F_D$ is the drag (lift forces are omitted). In general (from the Buckingham $\Pi$ theorem), the drag force can be expressed in the form
\begin{equation}
	F_D = \frac{1}{2} \varrho_L \mathbf u^2 A C_D,  
\end{equation}
where $A$ is the cross sectional area of the body, while $C_D$ is the drag coefficient -- a dimensionless number, which is dependent on the Reynolds number, see \cite{Batchelor 2000, Crowe et al. 2012} for details. Assuming that the considered body is sphere with the diameter $D$, the Reynolds number, defined as $Re=uD/\nu$, is low ($Re<1$ -- Stokes flow). Hence, the drag coefficient takes a very elegant form~\cite{Stokes 1851}
\begin{equation} \label{Stokes}
	C_D = \frac{24}{Re}.
\end{equation}
For higher Reynolds numbers (including the turbulent flow), there are many different propositions of correlations between the drag coefficient and the Reynolds number, see \cite{Crowe et al. 2012} for review (for non-spherical particles see \cite{Ouchene et al. 2015}). The most commonly used correlation which is reasonably for the Reynolds numbers up to $Re=800$ is the relationship proposed by \emph{Schiller and Naumann (1933)}~\cite{Schiller and Naumann 1933}
\begin{equation} \label{Schiller and Naumann}
	C_D = \frac{24}{Re}\left(1 + 0.15 Re^{0.687} \right).
\end{equation}
Assuming the Stokes regime of the flow, Eq.~(\ref{Stokes}), the solution of Eq.~(\ref{general equation of motion}) is
\begin{equation} \label{analytical velocity}
	u(t) = \frac{D^2g}{18\mu} (\varrho_S - \varrho_L) \left(1 - e^{-\frac{18 \mu}{D^2 \varrho_s} t} \right),
\end{equation}
which asymptotically approaches the terminal velocity
\begin{equation} \label{terminal velocity}
	u_T = \frac{D^2g}{18\mu} (\varrho_S - \varrho_L).
\end{equation}
The drag coefficient, for the terminal velocity, can be measured from the relationship
\begin{equation} \label{drag}
	C_D = \frac{4}{3} \frac{gD}{u_T^2} \left( \frac{\varrho_S}{\varrho_L} - 1 \right).
\end{equation}

It is important to note that Eqs.~(\ref{analytical velocity})-(\ref{drag}) explicitly depend on the diameter of the particle/droplet. This fact implies another question -- what is the diameter of the SPH particle seen by the fluid. Taking into account the initial homogeneous distribution of particles, the intuitive diameter of particle is $D=\Delta r$. On the other hand, in the SPH modeling we have two characteristic sizes which determine the resolution: $\Delta r$ and $h$. The third option is the particle diameter $D_{\Omega}$ determined from the assumption of particle sphericity and a constant volume $\Omega=m/\varrho$
\begin{equation}
	D_{\Omega} = \sqrt[3]{\frac{6}{\pi} \frac{m}{\varrho}} = \sqrt[3]{\frac{6}{\pi}} \Delta r.
\end{equation}
Due to this ambiguity, we decided to consider all three possibilities.

\begin{figure}
	\centering
	\begin{tabular}{cc}
    	\includegraphics[width=0.45\textwidth]{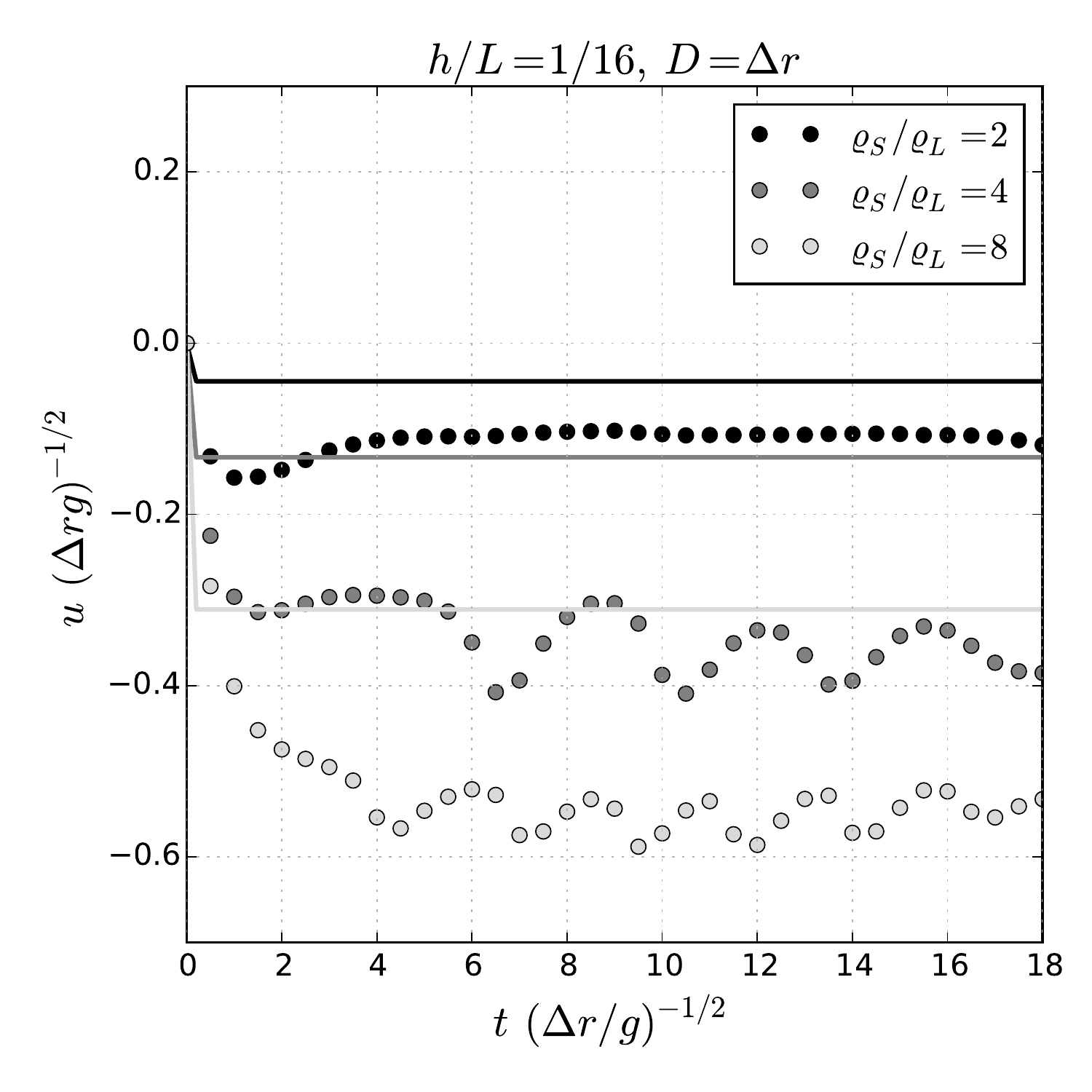} &
    	\includegraphics[width=0.45\textwidth]{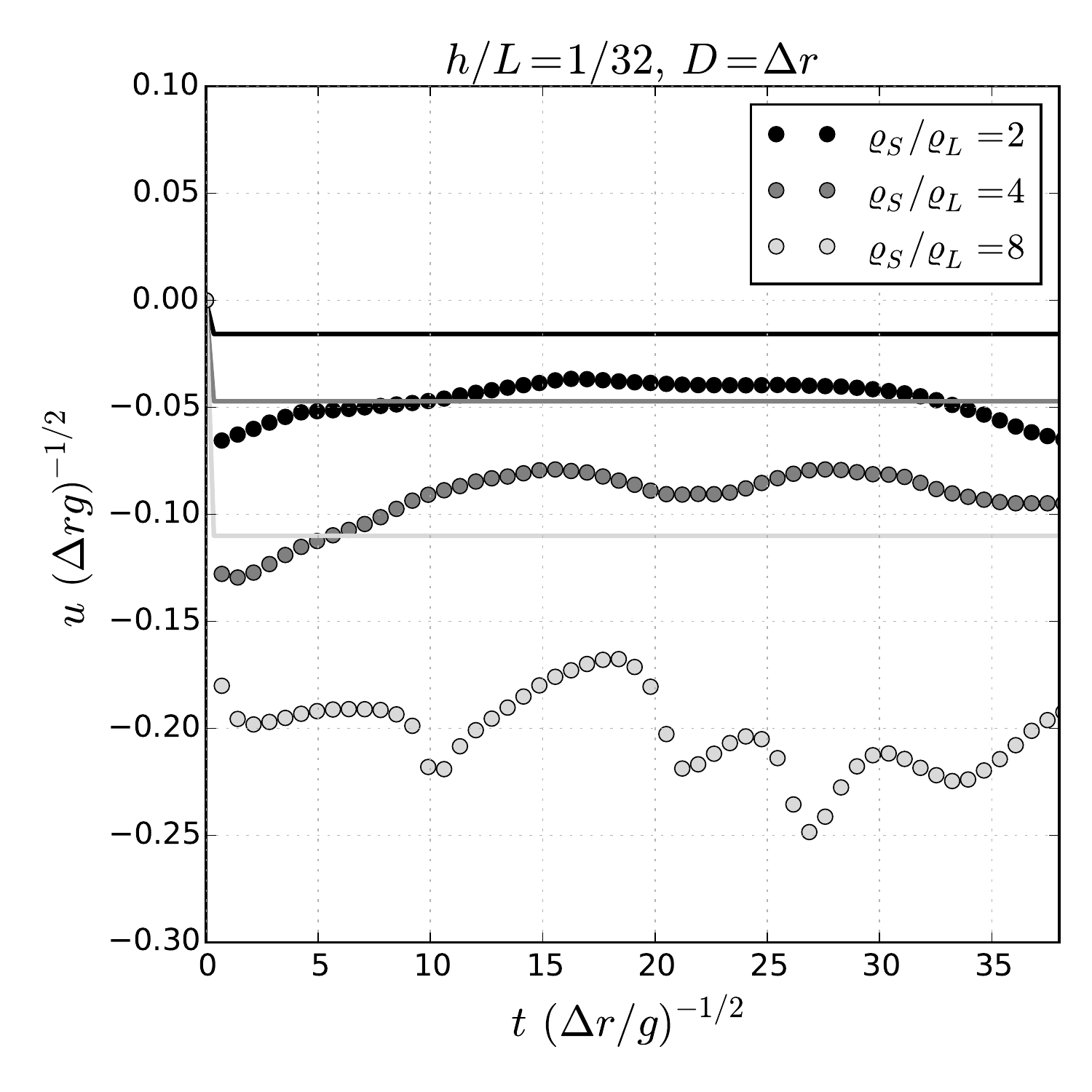} \\
    	\includegraphics[width=0.45\textwidth]{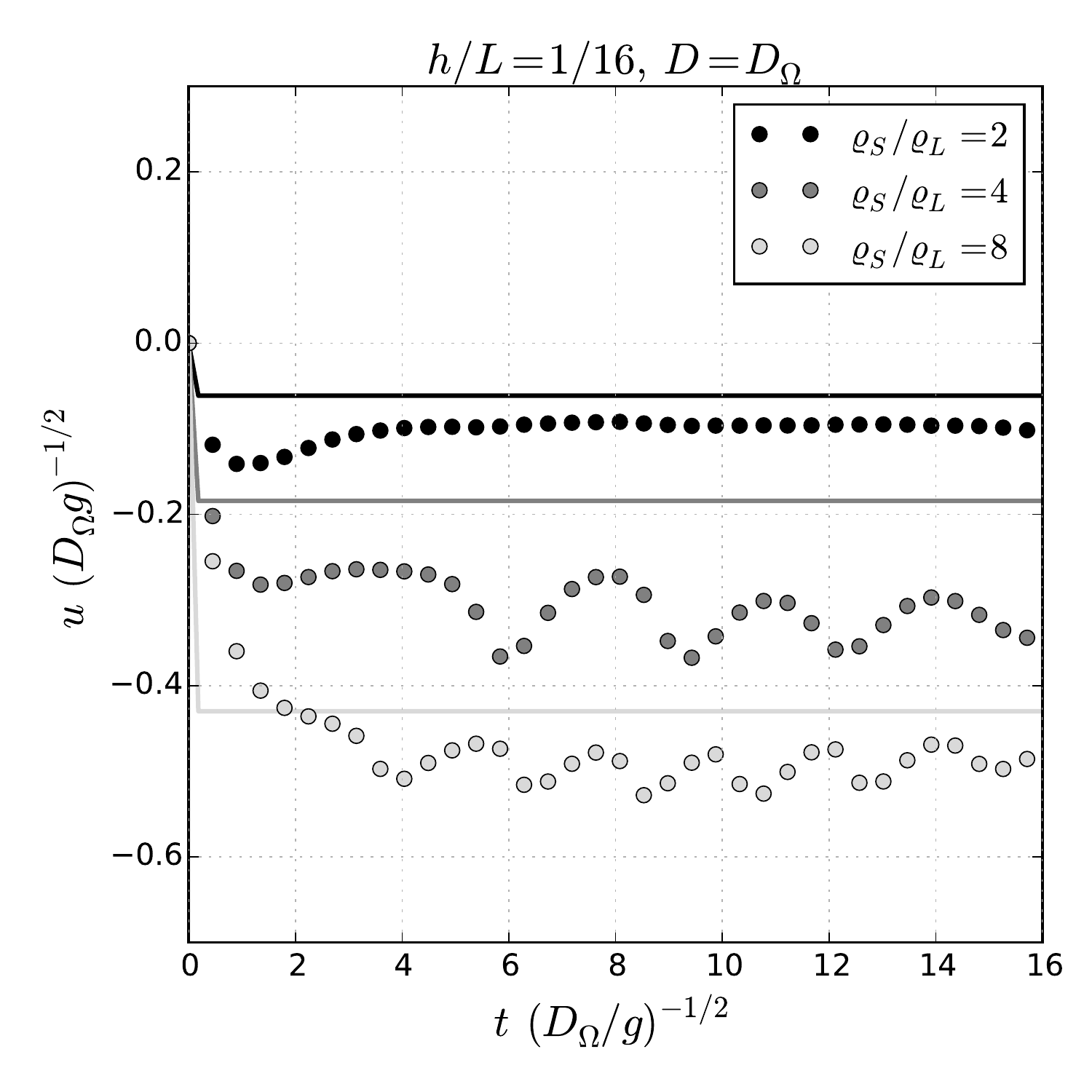} &
    	\includegraphics[width=0.45\textwidth]{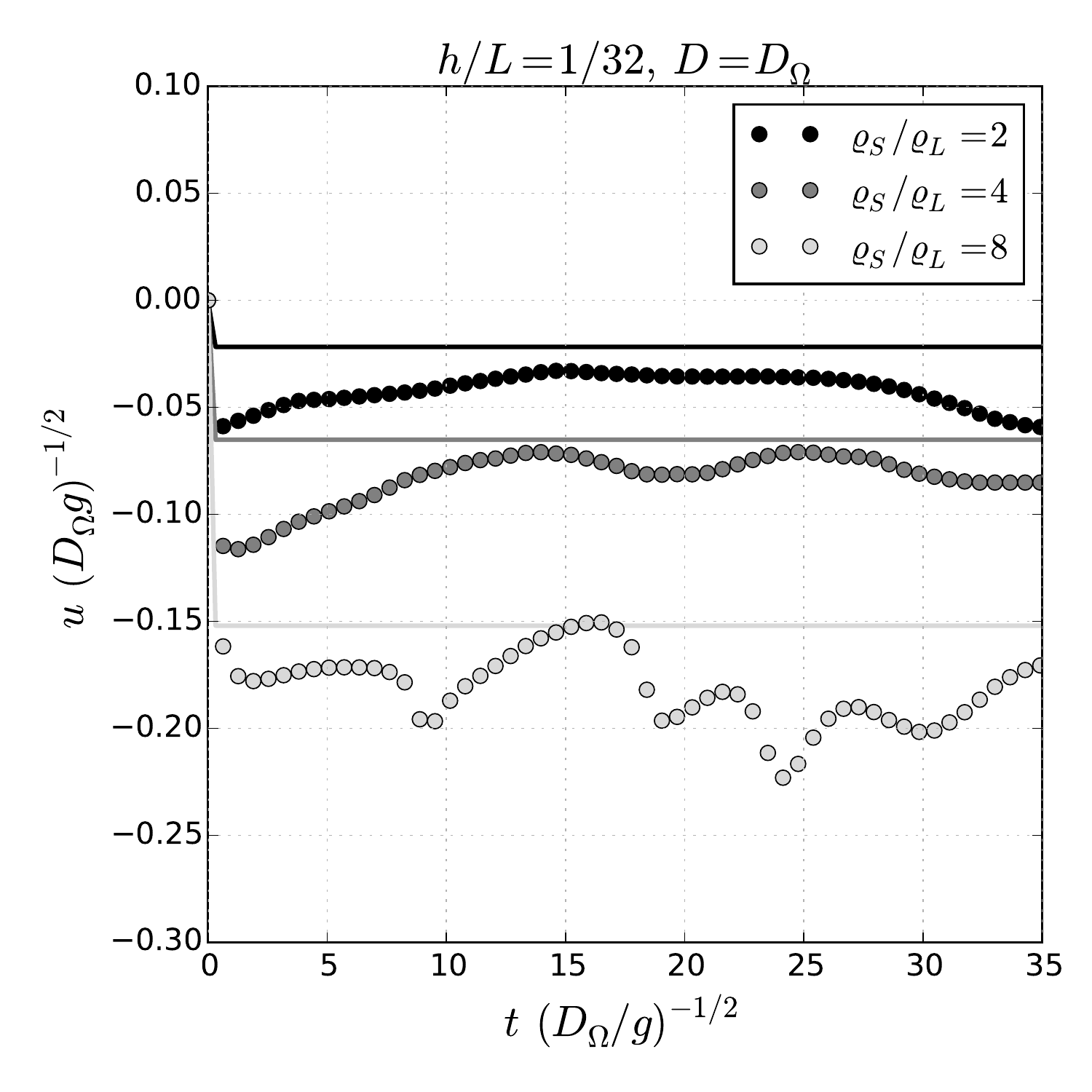} \\
    	\includegraphics[width=0.45\textwidth]{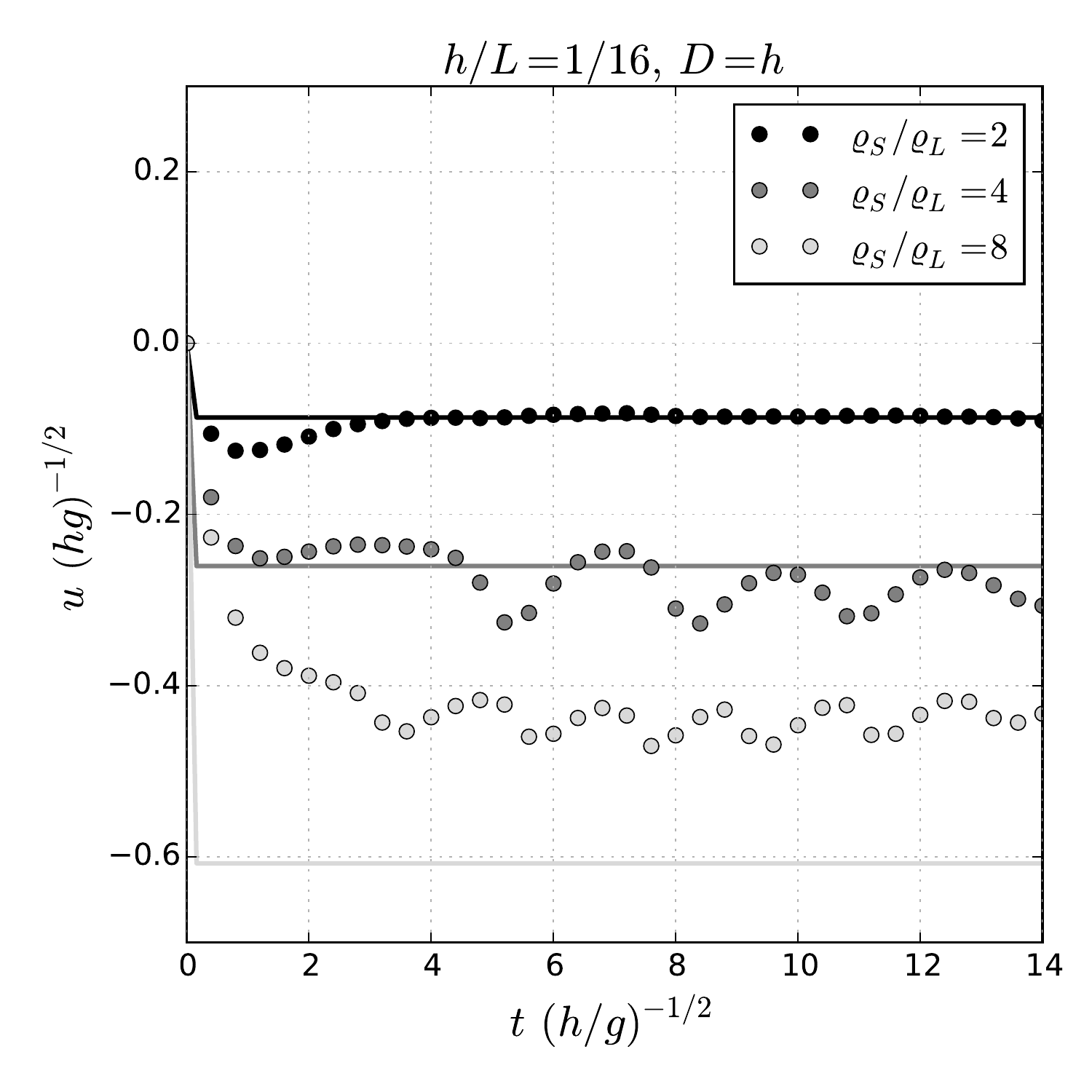} &
    	\includegraphics[width=0.45\textwidth]{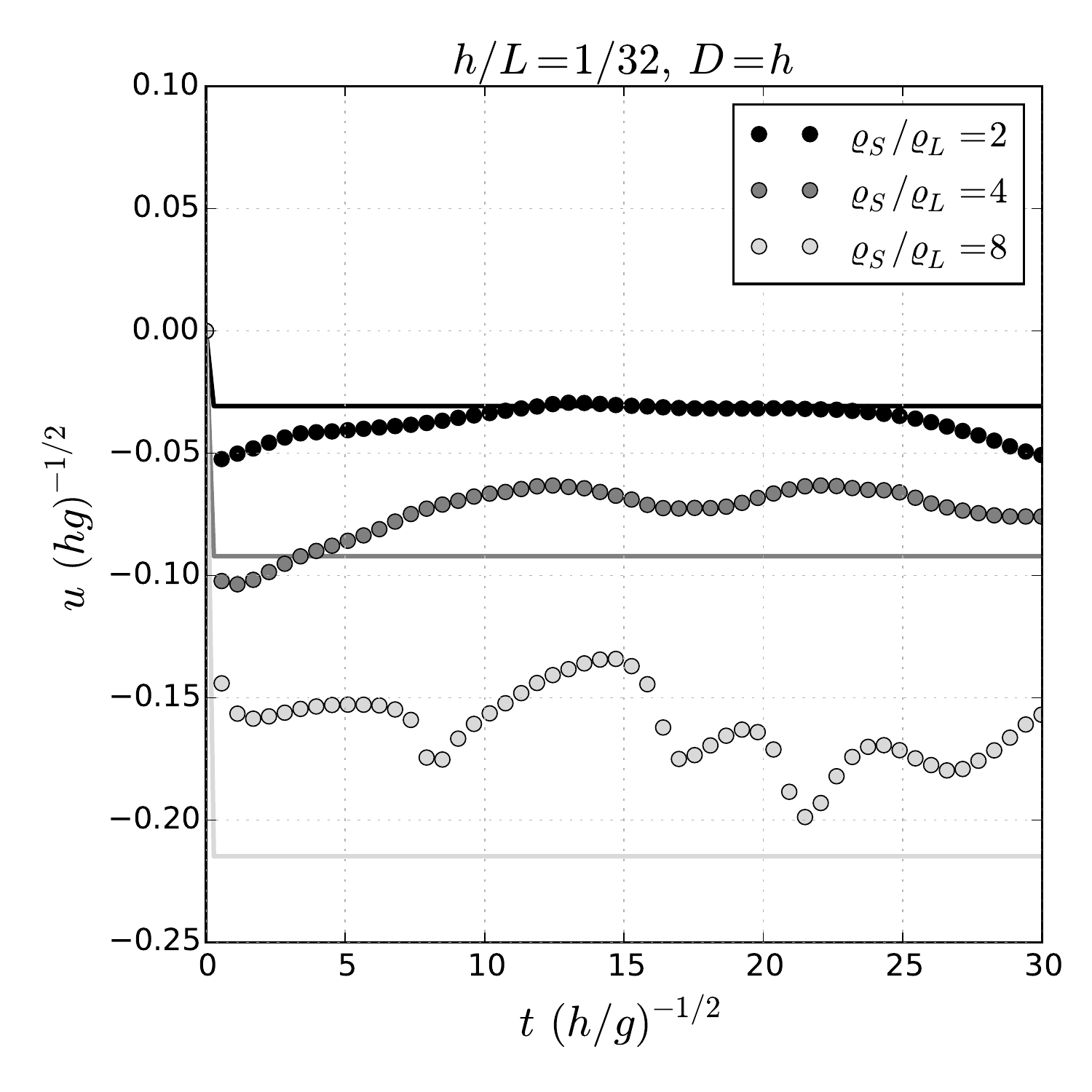}
    \end{tabular}
    \caption{Vertical velocity of solid particle falling down within fluid; the results obtained using the SPH approach (dots), and compared with the analytical predictions (solid lines), for different values of $h/L$ and different assumptions about the SPH particle diameter $D$. }
    \label{fig:v-t}
\end{figure}

In order to check whether the SPH method can give proper predictions of velocities when single particle/droplets are considered, we decided to perform a set of simulations for different density ratios, $\varrho_S/\varrho_L$ and two different resolutions: $h/L=1/16$ and $h/L=1/32$. In all simulations we decided to use the Wendland kernel and $h/\Delta r = 1.5625$. The dimensionless velocities calculated with the SPH method compared with the analytical predictions, Eq.~(\ref{analytical velocity}), are presented in Fig.~\ref{fig:v-t}. 
The obtained results show very high oscillations related to the interactions with the internal structure of the liquid, consisting of finite number of the SPH particles. However, the most interesting is a much better agreement with the reference solution if we assume that $D=h$. For $D=D_{\Omega}$ the results are slightly less accurate, while for $D=\Delta r$ we obtained the worst agreement with Eq.~(\ref{analytical velocity}).

\begin{figure}
    \centering
    \includegraphics[width=0.49\textwidth]{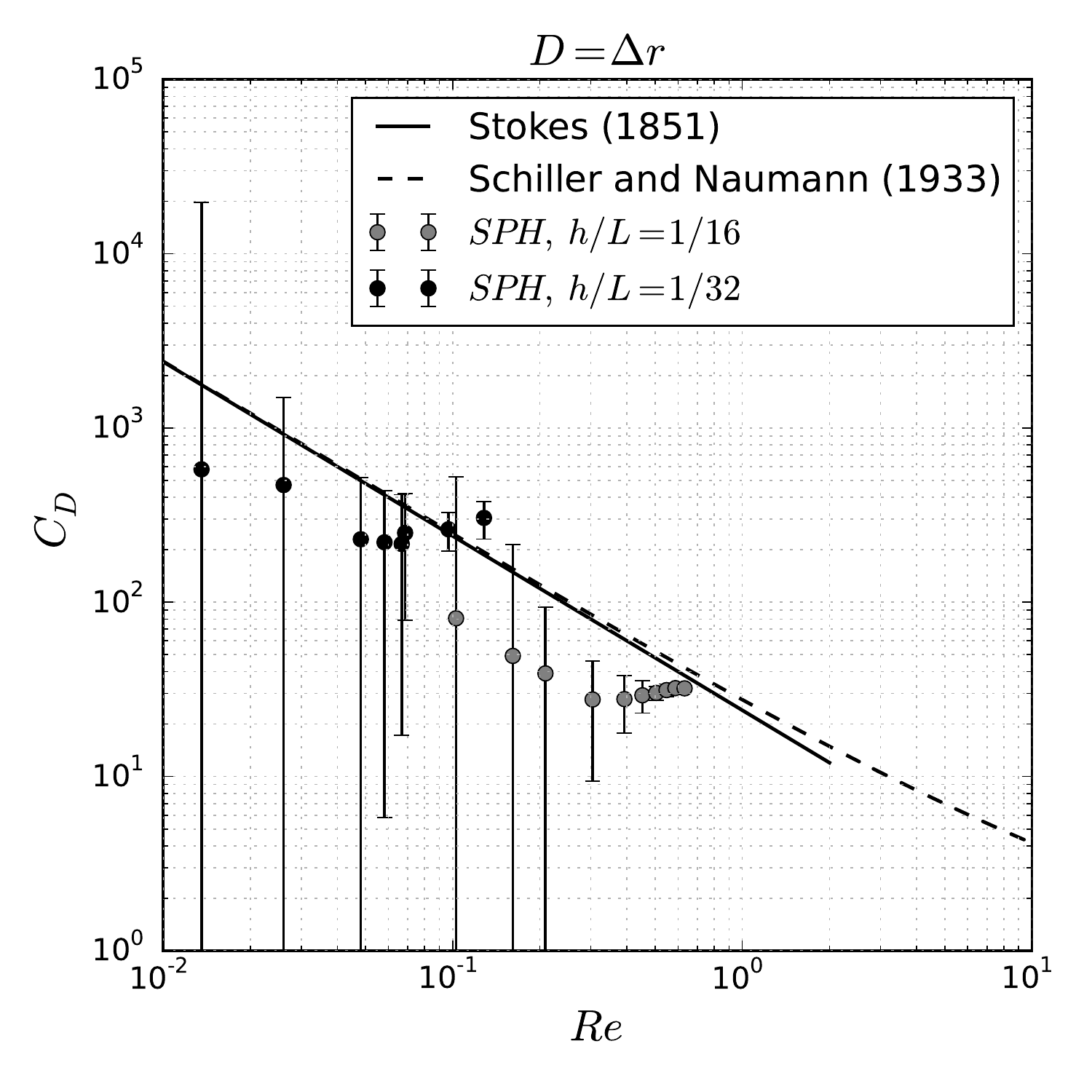}
    \includegraphics[width=0.49\textwidth]{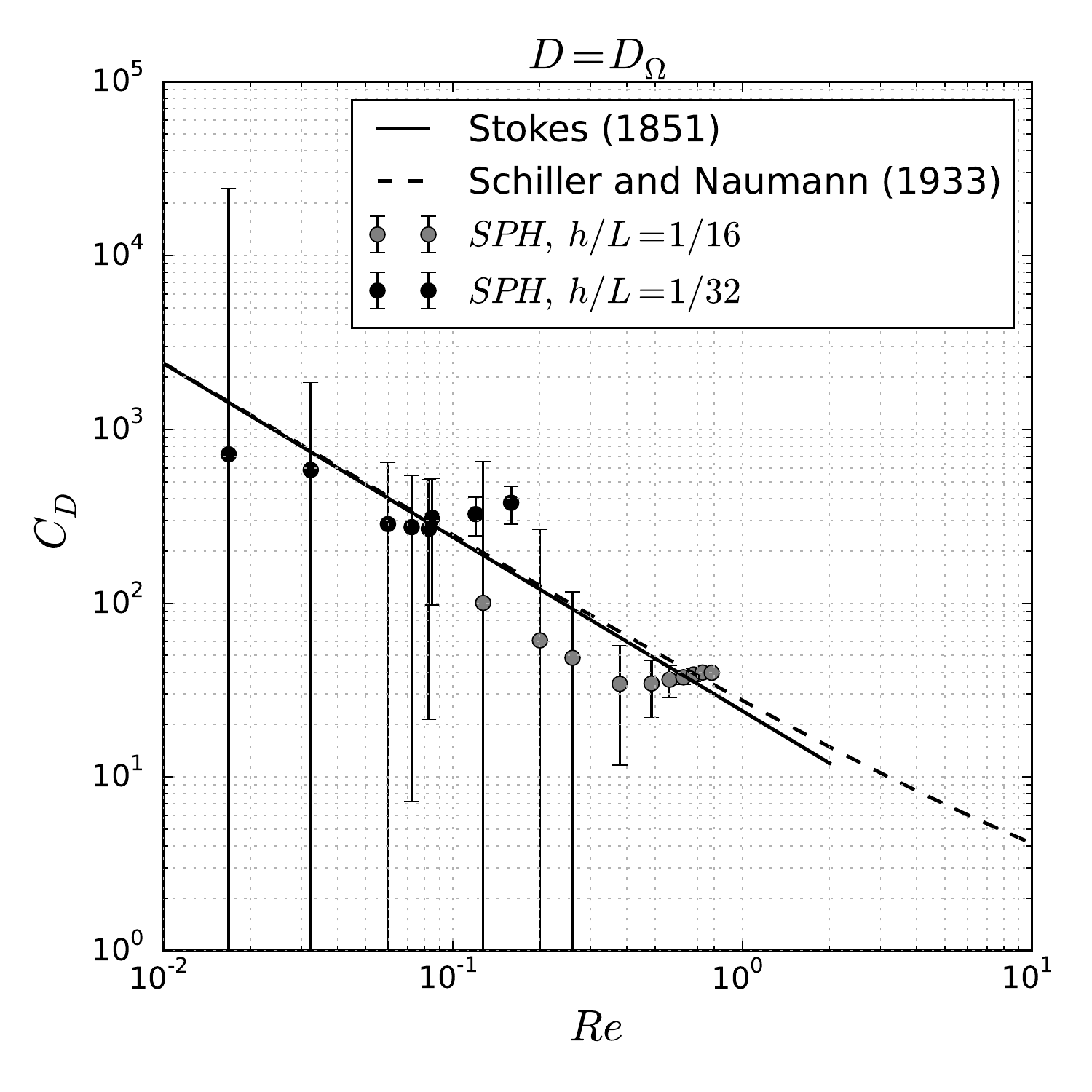}
    \includegraphics[width=0.49\textwidth]{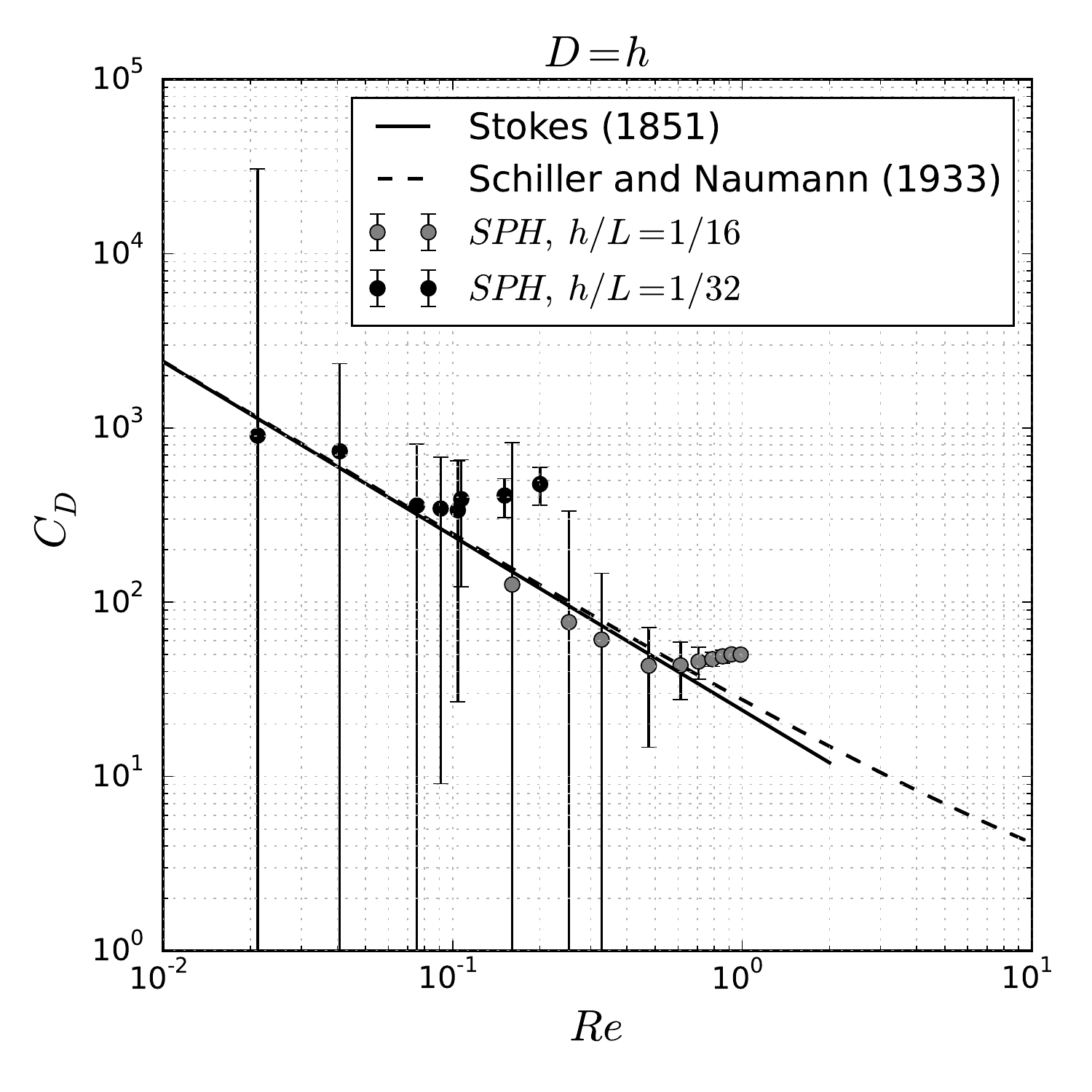} 
    \caption{Drag coefficient $C_D$ of falling particle as function of the Reynolds number $Re$ calculated using the SPH approach (dots); the SPH results are compared with the analytical (solid line) and experimental predictions (dashed line) for different assumptions about the particle diameter $D$. }
    \label{fig:cd-re}
\end{figure}

In the next step, we decided to calculate the drag coefficient $C_D$ for different density ratios which corresponds to different Reynolds numbers $Re$, and then compare it with the analytical \emph{Stokes (1851)}, Eq.~(\ref{Stokes}), and experimental \emph{Schiller and Naumann (1933)}, Eq.~(\ref{Schiller and Naumann}) correlations. The results obtained for two different resolutions: $h/L=1/16$ and $1/32$, and three different assumptions about the diameter of particle: $D=\Delta r$, $D_{\Omega}$ and $h$, are presented in Fig.~\ref{fig:cd-re}.
The first conclusion from these results is the underestimation of the drag coefficient in the case of $D=\Delta r$. The second conclusion is a very high standard deviation of the calculated drag coefficients for relatively slow particles, which is the result of the oscillations presented in Fig.~\ref{fig:v-t}. The third outcome is a high mismatch between the SPH calculations and the reference correlations for relatively fast particles (these with small errorbars). 

Therefore, taking into account the above-mentioned conclusions, the question about the physical correctness of the single particle droplets/solids should be answered: this motion is highly affected by SPH nodes (particles) and therefore non-physical. However, on the other hand, trying to be more \emph{open-minded}, since the results obtained for relatively low velocities (these with large errorbars in Fig.~\ref{fig:cd-re}) are with (high) margin of error in agreement with the reference data, we could potentially model large swarms of solid particles (e.g. sediments) interpreting the obtained results statistically.

However, very intriguing is the question about the diameter of the particle. The obtained results show much better agreement with the reference data for $D=h$ and $D=D_{\Omega}$, than $D=\Delta r$. Since the smoothing length can be chosen independently from the particle mass and density (for given $h$ mass is determined by the parameter $h/\Delta r$), a good agreement for $D=h$ is very surprising. In order to check whether this fact is just a coincidence, we decided to investigate the influence of the $h/\Delta r$ parameter on the resulting drag coefficient. The simulations were performed for three different values of $h/L=1/16$, $1/32$ and $1/64$, four different values of $h/\Delta r=1.5625$, $1.75$, $1.9375$, $2.125$ and constant $\varrho_S/\varrho_L=6$.
\begin{figure}
    \centering
    \includegraphics[width=0.55\textwidth]{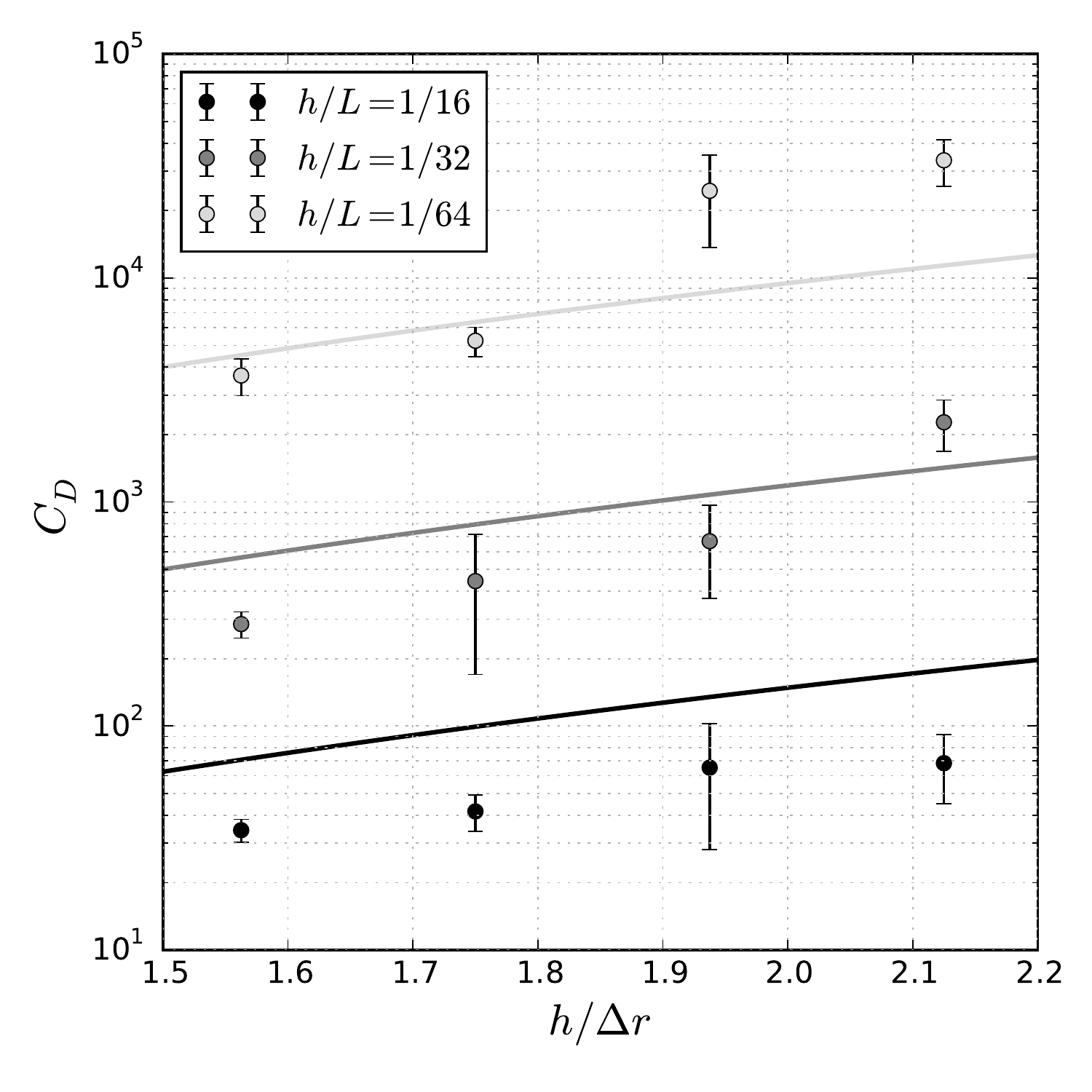} 
    \caption{Influence of $h/\Delta r$ parameter on the value of drag coefficient; comparison of results from simulations (dots) with the analytical predictions (lines), see Eq.~(\ref{re-hdr}).}
    \label{fig:cd-hdr}
\end{figure}
The obtained results are presented in Fig.~\ref{fig:cd-hdr}. The solid lines denote the expected, analytical (Stokes regime) dependency between the drag coefficient, $C_D$, and the parameter $h/\Delta r$
\begin{equation} \label{re-hdr}
	C_D = \frac{72 \mu^2 \pi}{g(\varrho_S-\varrho_L)\varrho_L h^3} \left( \frac{h}{\Delta r} \right)^3.
\end{equation}
Since both the SPH calculations and the analytical curve, Eq.~(\ref{re-hdr}), shows a similar upward trend with an increase in $h/\Delta r$, this indicates that the SPH particle diameter seen by fluid can not be $h$ -- in fact, it is $D_{\Omega}$.

\section{Conclusions}
Our simple numerical experiment yields important information on how single SPH particles should be viewed. The obtained results clearly indicate that movement of single SPH particles can not be considered as physically correct. Simulations of solid particles falling down in liquid showed that its behavior is highly affected by interactions with particles representing fluid. This issue is reflected in oscillations of velocity. Comparison of values of drag coefficient with the analytical and the experimental predictions also does not speak in favour of such a modeling using the SPH framework. While numerical results for relatively slow particles are within margin of error, agreement decreases significantly for faster ones. This leads towards conclusion, that in SPH simulations of multiphase flows, movement of single SPH particle solids/droplets should be regarded as numerical error, not psychical phenomenon. Furthermore ambiguities concerning size of single SPH particle has been cleared out. Among three candidates for diameter seen by fluid, i.e. $h$, $\Delta r$ and $D_{\Omega}$, the last one obtained from assumption of sphericity and constant volume $\Omega=m/\varrho$, was proved to be the best estimation.  

\section*{Acknowledgment} 
This research has been partly funded by the National Science Centre (Poland) via grant \emph{Opus 6} no DEC-2013/11/B/ST8/03818

\end{document}